\documentstyle [12pt] {article}
\begin{document}

\def\up{\uparrow}
\def\dw{\downarrow}
\def\ri{\rightarrow}
\def\inte{{_{\rm int}}}
\def\s{\hat{s}}
\def\a{{\cal A}_L}
\def\p{{\cal P}_L}
\def\s{\hat{s}}
\def\t{\hat{t}}
\def\u{\hat{u}}
\def\sig{\hat{\sigma}}
\def\half{{1\over 2}}
\def\non{\nonumber}
\def\np{{\sl Nucl. Phys.~}}
\def\pr{{\sl Phys. Rev.~}}
\def\pl{{\sl Phys. Lett.~}}

\def\va{\noalign{\vspace{4pt}}}

\font\el=cmbx10 scaled \magstep2

{\obeylines
\hfill ITP-SB-93-07
\hfill IP-ASTP-05-93
\hfill June, 1993}

\vskip 1.0 cm

\centerline {{\el Analysis of Weak-Interaction Effects in}}
\centerline {{\el High Energy Hadron-Hadron Collisions}}

\bigskip
\medskip

\centerline{\bf Hai-Yang Cheng$^{a,b}$, Melin Huang$^a$, and C. F. Wai$^a$}

\medskip
\centerline{$^a$ Institute of Physics, Academia Sinica,}
\centerline{Taipei, Taiwan 11529, Republic of China}

\medskip
\centerline{$^b$ Institute for Theoretical Physics, State University of New
York,}
\centerline{Stony Brook, NY 11794, USA}

\medskip
\bigskip
\medskip

\centerline{\bf Abstract}
\medskip

   Parity-violating (pv) effects in inclusive hadron and jet productions in
high energy hadron-hadron collisions are analyzed.
Such effects arise from the interference between strong and
weak amplitudes. This interference gives rise to a nonzero value of the pv
parameters $\a$ and $\p$, where $\a$ measures the difference in the inclusive
cross sections of, for example, $p+p\ri {\rm jet}+X$ ($X$=anything), with one
of incident proton beams in a state of
$\pm$ helicity, and $\p$ denotes the longitudinal polarization of a
high-energy baryon (e.g., $\Lambda$)
produced in $p+p\ri\Lambda+X$ with the initial proton beams unpolarized.
In the present paper, the single helicity asymmetry $\a$ in one-jet, two-jet
and two-jet plus photon productions as well as in the Drell-Yan process
$p+p\ri\ell^+\ell^-+{\rm jet}+X$ is probed, and the longitudinal polarization
$\p$ of the $\Lambda$ produced in unpolarized $pp$ collisions is studied.
We conclude that the pv effects in high energy proton-proton collisions are
in general only sensitive to the spin dependent valence quark distributions.

\vfill\eject

\noindent{\bf I.~~Introduction}
\vskip 0.5cm

  Hadron colliders with high-energy polarized proton beams are conceivably
available in the future at RHIC, SSC and LHC. Depending on whether the proton
beams are polarized
longitudinally or transversely, parton spin densities of the proton can be
probed via the studies of helicity or transverse spin asymmetries.
With longitudinal polarization, the double helicity asymmetry defined by
$${\cal A}_{LL}=\,{d\sigma^{++}-d\sigma^{+-}\over d\sigma^{++}+d\sigma^{+-}}
\eqno(1.1)$$
is the observable most commonly discussed in the literature, where $d\sigma
^{++}~(d\sigma^{+-})$ denotes the inclusive cross section for the
configuration where the incoming hadron's longitudinal polarizations are
parallel (antiparallel). Double asymmetries at high energies have been
investigated for various processes, such as single-jet [1], two-jet
[1,2], two-jet plus photon [3] and three-jet [3,4] productions, double-photon
production [5], direct photon production at large transverse momentum [6,7],
and the Drell-Yan process [7]. Most recent works were motivated by the
European Muon Collaboration
(EMC) measurement of the polarized proton structure function $g_1^p(x)$ [8].
The central issue of much theoretical controversy is whether or not gluons
contribute to the first moment of $g_1^p(x)$. Two extreme possibilities
for the explanation of the EMC experiment have been explored in the past:
large (negative) sea polarization [9] or large (positive) gluon polarization
[10]. Measurements of aforementioned processes will help determine the
spin dependent parton distributions and shed light on the interpretation
of the EMC results.

Contrary to the previous works, the purpose of the present paper is to analyze
the single helicity asymmetry $\a$ defined in Eq.(2.1) in high energy
proton-proton collisions. Experimentally, it should be easier to measure
$\a$ than the double helicity asymmetry. However, theoretically a nonzero $\a$
can occur only if some of the parton-parton scatterings involve
parity-violating weak interactions. Therefore, single helicity asymmetry
can be used to probe
parity violation in parton-parton subprocesses. Another party-violating
(pv) effect of interest is the longitudinal polarization $\p$ of a high-energy
baryon [see Eq.(2.2)] produced from unpolarized incident proton beams. Owing
to the small size of weak effects, pv parameters $\a$ and $\p$ arise from the
coherent interference between the strong-QCD and weak amplitudes. Such pv
effects were first analyzed in Ref.[11]
\footnote{This interference effect was also briefly discussed in Ref.[12].}
and subsequently in Ref.[13]. Specifically, the asymmetry parameter $\a$ for
the processes $p+p\to\pi^++X$ and $p+p\to{\rm jet}+X$, and the longitudinal
polarization of $\Lambda$'s in $p+p\to\Lambda+X$ were studied in Ref.[11]. The
content of the present work is in some sense the extension of the previous
analysis of Ref.[11].

  This paper is organized as follows. In Section II we discuss the general
formulism for calculating the pv parameters $\a$ and $\p$. It is stressed that
inspired by the EMC experiment and armed with the phenomenologically
determined valence-quark spin densities, two of us (H.Y.C. and
C.F.W.) have extracted the polarized sea and gluon distributions from the EMC
data for several different possibilities [14].
The pv effects in one-jet, two-jet, two-jet plus photon productions are
investigated in Sections III and IV.
\footnote{See Section VI for comments on the pv parameter $\a$ for $W^\pm$
and $Z^0$ productions in $pp$ collisions.}
The Drell-Yan process $p+p\to \ell^+\ell^-+{\rm jet}+X$ is studied in
Section V. The results are discussed in Section VI, where a series of figures
will be presented.

\vskip 0.6cm
\noindent{\bf II.~~General formalism}
\vskip 0.4cm

   In high energy proton-proton collisions, there are two single-helicity
asymmetry observables which we are interested in:
$$\a= \,{d\sigma^+-d\sigma^-\over d\sigma^++d\sigma^-},\eqno(2.1)$$
and
$$\p=\,{d\sigma_+-d\sigma_-\over d\sigma_++d\sigma_-}.\eqno(2.2)$$
In Eq.(2.1) $d\sigma^{\pm}$ denote the inclusive cross sections for $pp$
scattering where one of the initial proton beams is longitudinally polarized
and has $\pm$ helicity. In Eq.(2.2) $d\sigma_{\pm}$ denote the cross
sections for producing a high-energy baryon (e.g. $\Lambda$) in a state of
$\pm$ helicity from unpolarized proton beams. Both parameters are expected
to vanish to all orders in strong interactions. This can be easily seen
in the quark-parton model where the unpolarized inclusive differential cross
section for $pp$ collisions is given by
$$d\sigma=\,\sum_{i,j}\int dx_1dx_2\,f_i(x_1,~Q^2)f_j(x_2,
{}~Q^2)\left({d\hat{\sigma}_{ij}\over d\t}\right)d\t,\eqno(2.3)$$
where $f_i(x,~Q^2)$ is the unpolarized distribution
function of the parton $i$ in a proton with momentum fraction $x$,
 and $d\hat{\sigma}_{ij}$ is the
cross section for the interaction of two partons $i$ and $j$. When one
of the initial proton beams is longitudinally polarized, we have
$$d\sigma^+-d\sigma^-=\,\sum_{i,j}\int dx_1dx_2\Delta f_i(x_1,~Q^2)f_j(x_2,~
Q^2)\left[{d\hat{\sigma}\over d\t}(i^+j\to kl)-{d\hat{\sigma}\over d\t}(i^-j
\to kl)\right]d\t,\eqno(2.4)$$
with
$$\Delta f(x,~Q^2)=\,f_+(x,~Q^2)-f_-(x,~Q^2),~~~
f(x,~Q^2)=\,f_+(x,~Q^2)+f_-(x,~Q^2),\eqno(2.5)$$
where $f_\pm(x,~Q^2)$ is the parton distribution function in a polarized
proton with helicity parallel (antiparallel) to the proton spin, and ${d\hat{
\sigma}\over d\t}(i^\pm j\to kl)$ is the cross section for the scattering
$ij\to kl$
when parton $i$ has $\pm$ helicity. Since parity is conserved by strong
interactions, it is evident that $d\sigma^+=d\sigma^-$
and hence $\a=0$.

   If some of the parton-parton scattering subprocesses involve
parity-violating weak interactions, then in general
$d\hat{\sigma}(i^+j\to kl)\neq d\hat{\sigma}(i^-j\to kl)$ and thus a nonzero
$\a$ is expected. Owing to the small size of the weak effects, the
partiy-violating asymmetry $\a$ will arise from the coherent interference
between the strong-QCD amplitude, which is parity conserving, and the
parity-violating weak amplitude. Likewise, the longitudinal polarization $\p$
of a high-energy baryon, say $\Lambda$, produced in $p+p\to\Lambda+X$ with
unpolarized initial proton beams would also arise from the interference of
strong and weak amplitudes. Explicitly,
$$\begin{array}{lcl}
d\sigma_+-d\sigma_-= \,& \sum_{i,j,k,l} & \int dx_1dx_2dz\,f_i(x_1,~
Q^2)f_j(x_2,~Q^2)\\~~\\  & \times & \left[{d\hat{\sigma}\over d\t}(ij\to k^+l)
-{d\hat{\sigma}\over d\t}(ij\to k^-
l)\right]d\t\,\Delta D_k^\Lambda(z),  \end{array}\eqno(2.6)$$
with
$$\Delta D_k^\Lambda(z)=\,D_{k^+}^{\Lambda^+}(z)-D_{k^-}^{\Lambda^+}(z),
\eqno(2.7)$$
where $d\sigma_\pm$ is the differential cross section for producing parton $k$
with $\pm$ helicity, and $D_{k^\pm}^{\Lambda^+}(z)$ is the probability that
parton $k$ with $\pm$ helicity decays into $\Lambda$ with + helicity and
fractional momentum $z$.

   In order to estimate the partiy-violatin effects $\a$ and $\p$,
 we need  input of the polarized parton distribution functions
$\Delta f_i(x,~Q^2)$ and the polarized fragmentation functions $\Delta D
_q^\Lambda(z)$. Some useful information on the quark and gluon spin densities
can be obtained from the measurement of
the polarized proton structure function $g_1^p(x)$ in  deep
inelastic lepton-nucleon scattering.
Denoting the spin-dependent parton distributions by
$$\begin{array}{lcl}
\Delta q(x)& = &\,q^\up(x)+\bar{q}^\up(x)-q^\dw(x)-\bar{q}^\dw(x), \\
\Delta G(x)& = &\,G^\up(x)-G^\dw(x),  \end{array}\eqno(2.8)$$
we see that $\Delta q=\int^1_0\Delta q(x)dx$ and $\Delta G=\int^1_0\Delta G(x)
dx$ represent the net helicities carried by the quark flavor $q$ and the
gluon, respectively, in the infinite-momentum frame of the proton with
+ helicity. Since sea-quark and gluon polarizations are manifest
essentially in the region $x<0.1$, the valence-quark spin densities at
$x>0.1$ are constrained by the SLAC and EMC measurements of $g^p_1(x)$.
Following Ref.[14], the valence quark spin densities
parametrized at $Q_0^2=10\,{\rm GeV}^2$ are given by
$$\begin{array}{lcl}
\Delta u_v(x)& = &\,x^{0.287}u_v(x),   \\
\Delta d_v(x)& = &\,\left({x-x_0\over 1-x_0}\right)x^pd_v(x),   \end{array}
\eqno(2.9)$$
where $p=0.03,~0.26,~0.76$, for $x_0=0.35,~0.50$ and 0.75, respectively. For
unpolarized parton distribution functions we use the ``average'' set of
parametrization given in  Ref.[15] extracted also at the same reference
scale $Q^2_0=10\,{\rm GeV}^2$ from several experiments.
Eq.(2.9) follows from the perturbative QCD suggestion [16] that
valence quarks at $x=1$ remember the spin of the parent proton but become
totally unpolarized as $x$ is close to zero, and from the relations
$$\Delta u_v+\Delta d_v=\,3F-D,~~~\Delta u_v-\Delta d_v=\,F+D,\eqno(2.10)$$
with $F$ and $D$ being SU(3) parameters determined from neutron and
hyperon $\beta$ decays.
We find that this simple parametrization for $\Delta u_v(x)$ and $\Delta
d_v(x)$ fits to the data of $g^p_1(x)$ very well for $x>0.2$.

    In principle, the sea-quark polarization function $\Delta q_s(x)$ and the
gluon spin density $\Delta G(x)$ are constrained by the EMC measurement of
$g^p_1(x)$ at small $x$. However, the issue of whether or not gluons
contribute to the first moment of $g^p_1(x)$ has been under hot debate over
the past few years. Then it was realized that the size of hard-gluonic
contribution to $\int^1_0 g^p_1(x)dx$ is purely a matter of the factorization
convention chosen in defining the quark spin denisties $\Delta q(x)$
and the hard cross section for photon-gluon scattering $\Delta \sigma^{
\gamma G}(x)$ [17]. A change of the factorization scheme merely shifts the
contribution between $\Delta\sigma^{\gamma G}(x)$ and $\Delta q(x)$
in such a way that the polarized proton-photon cross section remains
unchanged. Depending on how the data of the polarized proton structure
function are explained, two of us (H.Y.C. and C.F.W.) have determined
$\Delta q_s(x)$ and $\Delta G(x)$ for three different possibilities from the
EMC measurement of $g^p_1(x)$ in conjunction with the above phenomenologically
determined valence-quark spin densities and the positivity constraint of
the unpolarized parton distributions [14]. In the case of the sea-quark
interpretation of the EMC experiment, we obtain
$${\rm case~1}:~~\cases{\Delta s(x)=\,-11.8\,x^{0.94}(1-x)^5s(x),  \cr
\Delta G(x)=0,  \cr}\eqno(2.11)$$
where we have assumed SU(3) invariance for sea polarizations $\Delta u_s(x)=
\Delta d_s(x)=\Delta s(x)$.
On the contrary, if the data of $g^p_1(x)$ are explained in terms of polarized
gluon and valence-quark densities, we find that an acceptable $\Delta G(x)$
does exist for properly chosen prescription; more precisely,
$${\rm case~2}:~~\cases{\Delta s(x)=0,  \cr
\Delta G(x)=6.0\,x^{0.76}(1-x)^3G(x).  \cr}\eqno(2.12)$$
Since in a realistic case, it is unlikely that $\Delta G(x,\,Q^2)$ or $\Delta
s(x,\,Q^2)$ vanishes at some scale $Q^2_0$ for all $x$,  it is more
pertinent to consider the case with non-vanishing $\Delta G(x)$ and $\Delta
s(x)$. In Ref.[14]
we have proposed more realistic spin-dependent sea and gluon
distribution functions which are parametrized in such a way that the first
moment of $g^p_1(x)$ receives almost all contributions from the region
$x>0.01$, as indicated by the EMC experiment:
$${\rm case~3}:~~\cases{\Delta s(x)=\,-3.39\,x^{0.62}(1-x)^{1.4}s(x),  \cr
\Delta G(x)=\,2.69\,x^{0.76}(1-x)^3G(x).   \cr}\eqno(2.13)$$
Case 3 is between the sea-quark (case 1) and gluon (case 2) interpretation of
the EMC
data. The above three sets of polarized sea and gluon distribution
functions are all parametrized at the reference scale $Q^2_0=10\,{\rm GeV}^2$.
Their $Q^2$ evoluations are obtained by solving the Altarelli-Parisi equation
numerically [18].

 Very little is known about the fragmentation function $D_q^\Lambda(z)$. Just
as the case of parton distribution functions, we may assume that $\Delta D_q^
\Lambda(z)$ is proportional to the unpolarized fragmentation function
$D_q^\Lambda(z)$. For these we take the simple parametrization [19]
$$zD^\Lambda_q(z)\equiv z D_u^\Lambda(z)=zD^\Lambda_{d}(z) = z D_s^\Lambda(z)
= B_v z(c-z)^3+B_s(1-z)^4,
\eqno(2.14)$$
and
$$zD^\Lambda_{\bar{q}}(z)\equiv zD^\Lambda_{\bar{u}}(z)=zD^\Lambda_{\bar{d}}
(z)=zD^\Lambda_{\bar{s}}(z)=B_s(1-z)^4.\eqno(2.15)$$
Assuming $D^\Lambda_q(z)\approx D^p_q(z)$, we find that a fit to the EMC data
of $D^p_u(z)$ [20] yields
$$c=\,1.08\,,~~B_v=\,0.5\,,~~B_s=\,-0.06\,.\eqno(2.16)$$

  For the polarized fragmentation functions, the construction of $\Delta D_q
^\Lambda(z)$ is necessarily somewhat {\it ad hoc} in the absence of
experimental data or a detailed theory. In analog to the quark distributions,
we simply assume that
$$\Delta D^\Lambda_q(z)=\,z^\gamma\,D^\Lambda_q(z).\eqno(2.17)$$
That is, the polarization of the outgoing $\Lambda$ is equal to that of the
parent quark at $z=1$ but diminishes as $z\to 0$. Recently it has been pointed
out by Burkardt and Jaffe [21] that a measurement of the helicity asymmetric
cross section for semi-inclusive production of $\Lambda$ in $e^+e^-$
annihilation near the $Z^0$ resonance allows a complete determination of
$\Delta D^\Lambda_q(z)$. This experiment should be practical at the LEP
collider at CERN or at SLC at SLAC.

\vskip 0.6 cm
\noindent{\bf III.~~Hadronic jet and $\Lambda$ productions}
\vskip 0.4 cm

    In this section we shall study the asymmetry parameter $\a$ for $pp\to{\rm
jet}+X$ and $pp\to{\rm 2~jets}+X$ and the longitudinal polarization $\p$
for $pp\to
\Lambda+X$. Hadronic jet productions are expected to be the main processes
in high-transverse-momentum hadron-hadron collisions. At the parton
level the parton-parton cross sections are the same for 1-jet, 2-jet and
$\Lambda$ productions, so we put them all together in this section.

   As emphasized in passing, we are interested in the differential cross
section $d\hat{\sigma}/d\hat{t}(ij\to kl)$ arising from the coherent
interference
between the strong QCD amplitude and the weak amplitude. In order for there to
exist any interference, it is necessary that these amplitudes connect the same
initial and final states. Since the strong QCD interactions change color but
conserve flavor, while weak couplings always conserve color, only a limited
number of quark and gluon scattering processes can give rise to a strong-weak
interference. Evidently, the external gluons in general cannot make
contributions to $\a$ and $\p$ at
tree level. The strong quark-quark scattering processes can be classified
into five categories: $qq\to qq,~qq'\to qq'~(q'\neq q),~q\bar{q}\to q\bar{q}$,
$q\bar{q}\to q'\bar{q}',~q\bar{q}'\to q\bar{q}'$. The differential cross
sections
for quark-quark scattering due to QCD and weak interference with one of the
initial quarks polarized are given in Ref.[11]. For our purposes,
it is more convenient to present these cross sections for both initial
quarks or antiquarks having definite helicities.
The color-averaged differential cross section due to the interference
between the strong and weak amplitudes is given by
\renewcommand{\theequation}{3.\arabic{equation}}
\begin{eqnarray}
{d\hat{\sigma}_{\rm int}\over d\t} & = & {1\over
32\pi\hat{s}^2}\left({4\over 9}
\right)|M(\lambda_1,\lambda_2,\lambda_3,\lambda_4)|^2, \nonumber \\
   & = & {16\over 9}{G_F\over \sqrt{2}}\alpha_s|T(\lambda_1,\lambda_2,
\lambda_3,\lambda_4)|^2,  \end{eqnarray}
for the quark-quark scattering $q_1(k_1,\,\lambda_1)q_2(
k_2,\,\lambda_2)\to q_3(k_3,\,\lambda_3)q_4(k_4,\,\lambda_4)$, where ${4\over
9}$ is a color-averaged factor. The results are summarized in Table I in four
different helicity states denoted by
\begin{equation}
\left( \matrix{ q_{_{1L}}q_{_{2L}} & q_{_{1L}}q_{_{2R}}  \cr  q_{_{1R}}q_{_{
2L}} & q_{_{1R}}q_{_{2R}} \cr}\right).
\end{equation}
Note that in Table I,
\begin{equation}
L_q=\,2T_3-2e_q\sin^2\theta_W,~~~R_q=\,-2e_q\sin^2\theta_W,
\end{equation}
are the left-handed and right-handed coupling constants respectively of the
quark coupled
with the $Z$ boson, where $T_3$ is an SU(2) isospin quantum number, $e_q$ is
the charge of the quark, and $\theta_W$ is the Weinberg angle. In Eq.(3.1)
$\s,~\t$ and $\u$ are the usual Mandelstam variables
\begin{equation}
\s=\,(k_1+k_2)^2,~~\t=(k_1+k_3)^2,~~\u=\,(k_2+k_3)^2,
\end{equation}
with $k_1+k_2+k_3+k_4=0$, $\Gamma_W~(\Gamma_Z)$ is the decay width of the
$W$ ($Z$) boson,
and $V_{qq'}$ is a quark mixing matrix element. It should be stressed that our
 results for $d
\hat{\sigma}_\inte/d\t$ are in agreement with that first derived in Ref.[11]
but differ from Ref.[13] in signs for some of the subprocesses.

\begin{table*}[t]
\begin{center}
\begin{tabular}{lccccccc}
\multicolumn{8}{l}{Table~I. The differential cross section $d\hat{\sigma}_
\inte/d\t$ for various quark-quark} \\
\multicolumn{8}{l}{\hbox to 48 pt{\hfil}scattering processes due to the
coherent interference between strong} \\
\multicolumn{8}{l}{\hbox to 48 pt{\hfil}and
weak amplitudes. Shown are the matrix elements absolute squared} \\
\multicolumn{8}{l}{\hbox to 48 pt{\hfil}$|T|^2$ [see Eq.(3.1)] in four
different helicity states denoted by (3.2).}  \\
\hline\hline
process & & $|T|^2$  \\
\noalign{\vspace{4pt}}
\hline
\noalign{\vspace{4pt}}
$qq\to qq$ && $\left[{M_Z^2\over \t(\u-M_Z^2)}+{M_Z^2\over \u(\t-M_Z^2)}\right]
\left(\matrix{ L_q^2 & 0 \cr 0 & R_q^2  \cr}\right)$  \\
\va
$qq'\to qq'$& & $-{8M_W^2\over \t(\u-M_W^2)}|V_{qq'}|^2\left(\matrix{1 & 0 \cr
0 & 0 \cr}\right)$ \\
\va
$q\bar{q}\to q\bar{q}$& & ${\u^2\over\s^2}\left[{(\s-M_Z^2)M_Z^2\over \t[(\s-
M_Z^2)^2+
\Gamma_Z^2M_Z^2]}+{M_Z^2\over \s(\t-M_Z^2)}\right]\left(\matrix{ 0 & L_q^2  \cr
R_q^2 & 0  \cr}\right)$  \\
\va
$q\bar{q}\to q'\bar{q}'$ &&  $-{8\u^2\over \s^2}{M_W^2\over \s(\t-
M_W^2)}|V_{qq'}|^2\left(\matrix{ 0 & 1  \cr  0 & 0  \cr}\right)$   \\
\va
$q\bar{q}'\to q\bar{q}'$ && $-{8\u^2\over \s^2}{(\s-M_W^2)M_W^2\over \t[(\s-
M_W^2)^2+
\Gamma_W^2M_W^2]}|V_{qq'}|^2\left(\matrix{ 0 & 1 \cr 0 & 0  \cr}\right)$ \\
\va
\hline\hline
\end{tabular}
\end{center}
\end{table*}

   The weak amplitudes for the processes $qq'\to qq',~q\bar{q}\to q'\bar{q}'$
and $q\bar{q}'\to q\bar{q}'$ receive both charged- and neutral-current
contributions. However, it is clear from Table I that only the weak amplitudes
due to $W$ exchange interfer with the strong amplitude. Since only left-handed
quarks and right-handed antiquarks participate in weak couplings with the $W$
boson, this explains why there is only one non-vanishing matrix element for
above-mentioned quark-quark scattering processes.

  It follows from Eq.(2.4) that the differential cross section for a single
jet production in $pp$ collisions at transverse momentum $p_T$ and rapidity
$y$ is given by
\begin{eqnarray}
E{d\sigma^+\over d^3p}-E{d\sigma^-\over d^3p} & = & \,{1\over \pi}\sum_{i,j}
\int_{x_0}^1 dx_a{x_ax_b\over 2x_a-x_Te^y}\Delta f_i(x_a,\,Q^2)f_j(x_b,\,Q^2)
\non \\ & \times & \left[{d\hat{\sigma}\over d\t}(i^+j\to kl)-{d\hat{\sigma}
\over d\t}(i^-j\to kl)\right],  \end{eqnarray}
where
\begin{equation}
x_T={2p_T\over \sqrt{s}},~~x_0=\,{x_Te^y\over 2-x_Te^{-y}},~~x_b=\,{x_ax_T
e^{-y}\over 2x_a-x_Te^y},\end{equation}
and it is to be understood that
\begin{eqnarray}
{d\hat{\sigma}\over d\t}(i^+j\to kl)-{d\hat{\sigma}\over d\t}(i^-j\to kl)=
\,& {1\over 2} & \bigg[  {d\sig\over d\t}(i^+j^+\to kl)+{d\sig\over d\t}(i^+
j^-\to kl) \non \\   & - & {d\sig\over d\t}(i^-j^+\to kl)-{d\sig\over d\t}(i^-
j^-\to kl)\bigg].    \end{eqnarray}
For the production of two jets with rapidities $y_1$ and $y_2$, and with
equal and opposite transverse momentum $p_T$, the differential cross section
reads [22]
\begin{eqnarray}
{d\sigma^+\over dy_1dy_2dp^2_T}-{d\sigma^-\over dy_1dy_2dp^2_T} & = & \,x_a
x_b\sum_{i,j}\Delta f_i(x_a,~Q^2)f_j(x_b,~Q^2)\non  \\  & \times & \left[
{d\sig\over d\t}(i^+j\to kl)-{d\sig\over d\t}(i^-j\to kl)\right],
\end{eqnarray}
where
\begin{equation}
x_a=\,{1\over 2}x_T\left(\tan^{-1}{\theta_1\over 2}+\tan^{-1}{\theta_2\over
2}
\right),~~~x_b=\,{1\over 2}x_T\left(\tan{\theta_1\over 2}+\tan{\theta_2\over 2}
\right), \end{equation}
with $\theta_1$ ($\theta_2$) being the angle between the first (second) jet
and one of the incident proton beams (see Ref.[23] for detail), and $\theta_1
+\theta_2=180^\circ$.
As for the longitudinal polarization $\p$ of $\Lambda$ produced in $pp\to
\Lambda+X$, the numerator of Eq.(2.2) is given by [23]
\begin{eqnarray}
E{d\sigma_+\over d^3p}-E{d\sigma_-\over d^3p} & = & {1\over \pi}\sum_{i,j,k,l}
\int^1_{x_a^{\rm min}}dx_a\int^1_{x_b^{\rm min}}dx_bf_i(x_a,~Q^2)f_j(x_b,~Q^2)
\non \\  & \times & \left[ {d\sig\over
d\t}(ij\to k^+l)-{d\sig\over d\t}(ij\to k^-l)\right]{1\over z}\Delta D^\Lambda
_k(z),  \end{eqnarray}
where
\begin{equation}
x_a^{\rm min}=\,-{u\over s+t},~~~x_b^{\rm min}=\,-{tx_a\over u+sx_a},~~~z=-{t
\over sx_b}-{u\over sx_a}.
\end{equation}

     The denominator of (2.1) and (2.2) is twice the unpolarized cross section,
whose explicit expression is not written here. The general features of
unpolarized cross sections
 are known. Since gluon densities are large at small $x$, gluon-gluon
scattering dominates the underlying parton-parton interaction
subprocesses at small $x_T$. As the jet momentum increases, quark-gluon
scattering becomes more and more important due to the relatively fast
decrease of the gluon distribution with increasing $x$. It is finally
governed by quark-quark scattering at large $x_T$.

\vskip 0.6 cm
\noindent {\bf IV.~~~Two-jet plus photon production }
\vskip 0.4 cm

      At the parton level,  the production of 2-jet plus photon final state
 at large transverse momentum can proceed via processes (denoted by $H_i$)
where a photon is emitted from a quark or antiquark, and processes
(denoted by $W_i$) where a pair $V-\gamma$ is produced with $V(=W^\pm~{\rm
or}~Z^0)$ decaying hadronically.  The amplitudes
for $H_i$ and all other amplitudes for 2-2 parton scattering involving
at least one quark or anti-quark off which $V$ is radiated,  give rise to the
dominant QCD background.  The amplitudes for $W_i$ with $V$ being far off
shell yield a much smaller cross section of order $\alpha^3$, compared with
the amplitudes $H_i$, whose corresponding cross sections are of order $\alpha
 \alpha^2_s$.  However, the former cross section can be possibly greatly
enhanced if $V$ is on its mass shell. This can be seen by comparing the boson
propagator $i/( k^2 - M^2_V + iM_V \Gamma_V)$
 with that of a massless quark or gluon $i/(k^2 + i\epsilon)$ with
roughly the same momentum $k$. Since ${M_V}^2/{\Gamma_V}^2 \cong 1.4
\times 10^3$ for $V=W^\pm$ or $Z^0$, it should help this production
cross section rise above the QCD background.

The tree-level processes which give rise to two-jet plus one real photon
production
events, have five possible classes, $qq \rightarrow qq\gamma$, $qq' \rightarrow
   qq'\gamma$,
$q\bar q \rightarrow q\bar q\gamma$, $q\bar q\rightarrow q'\bar q'\gamma$ and
$q\bar q'\rightarrow q\bar q'\gamma$.  We shall only write down those
interference terms involving QCD and electroweak interactions.

For the case of the subprocess $q(k_1)q(k_2)\to q(k_3)q(k_4)\gamma(k_5)$, the
absolute value squared of
spin dependent matrix-elements is given by
\renewcommand{\theequation}{4.\arabic{equation}}
\setcounter{equation}{0}
\begin{eqnarray}
\vert M(++)\vert ^2  & = & {\s^2_{12}+\s^2_{34} \over \s_{45} }e_q^2R_q^2
      \Bigg\{ (f_3 - g_2) \left[{1 \over {\s_{23}(\s_{24}-M_Z^2)}}
               + {1 \over {\s_{24}(\s_{23}-M_Z^2)}}\right] \non \\
      & - &  g_3 \left[{1\over {\s_{23}(\s_{13}-M_Z^2)}} + {1 \over{\s_{13}(
\s_{23}-M_Z^2)}}\right]   \non \\
      & + & (- g_1 + g_2 +g_3-f_1)  \left[{1 \over {\s_{14} (\s_{24} - M_Z^2)
}} +{1\over {\s_{24}(\s_{14}-M_Z^2)}}\right] \non  \\
      & - & (f_1 + g_1) \left[ {1 \over {\s_{14}(\s_{13} - M_Z^2)}} + { 1
\over {\s_{13} (\s_{14} - M_Z^2)}}\right] \Bigg\},
\end{eqnarray}
where $\s_{ij}=(k_i+k_j)^2$, and
\begin{eqnarray}
f_1 & = & 2{\s_{12} \over \s_{35}},\qquad f_2 = 2{\s_{13} \over \s_{25}},
\qquad f_3 = 2{\s_{23} \over \s_{15}},  \non \\
g_1 & = & {{\s_{25}\s_{13} +\s_{35}\s_{12}-\s_{23}\s_{15}} \over{\s_{25}
\s_{35}}},
\qquad g_2 = {{\s_{25}\s_{13} - \s_{15}\s_{23} - \s_{35}\s_{12}}\over
{\s_{35}\s_{15}}},      \non  \\
g_3 & = &{\s_{15}\s_{23}+ \s_{25}\s_{13} -\s_{12}\s_{35}\over {\s_{15}\s_{25}
}},     \end{eqnarray}
and $|M(--)|^2$ is obtained from $|M(++)|^2$ with the replacement $R_q\to L_q$.
For the case of $qq' \rightarrow qq' \gamma $ scattering, the non-vanishing
matrix element absolute squared is
\begin{eqnarray}
\vert M(--) \vert ^2 &= & {{\s^2_{12}} \over \s_{45}}
 \bigg\{ -{e_q \over \s_{23}} ( A g_2
 + B g_3 + C f_3 ) \non \\
&  + & {e_{q'} \over \s_{14}} [A(f_1 + g_1) +
                   B( f_2 + g_1 ) + C (g_2 + g_3)]\bigg\}|V_{qq'}|^2 \non \\
       &  + & {\s^2_{34} \over \s_{45}}\bigg\{ {e_{q'}\over \s_{13}-M^2_W}
\left[ {e_q \over \s_{23}} g_2 - {e_{q'} \over \s_{14}} (f_1 + g_1)\right]
\non  \\
& + & B \left[{e_q \over\s_{23}} g_3 - {e_{q'}\over\s_{14}} (g_1 + f_2)\right]
- C \left[ {e_q \over \s_{23}} f_3 - {e_{q'} \over \s_{14}} (g_2 + g_3)\right]
                 \bigg \}|V_{qq'}|^2, \end{eqnarray}
where
\begin{eqnarray}
A & = & {\s_{35}(\s_{25}+\s_{45})\over\s_{12}(\s_{13}-M^2_{_W})(\s_{24}-M^2_
{_W})}+ {\s_{35} \over \s_{12}(\s_{13}-M^2_{_W})}   \non  \\
   &   & - \left( {\s_{34}+\s_{45} \over \s_{12}}e_{q'} + e_q\right)
     {1 \over \s_{24}-M^2_{_W}},  \non \\
B & = & {e_{q'}\over\s_{13}-M^2_{_W}}+{\s_{25}\over{(\s_{13}-M^2_{_W})
(\s_{24}-M^2_{_W})}}, \non  \\
C & = & {e_q \over \s_{24}-M^2_{_W}}. \end{eqnarray}
For $\bar{q}\bar{q}'\to\bar{q}\bar{q}'\gamma$ scattering, the nonvanishing
matrix element squared is $|M(++)|^2$, which has the same expression as
Eq.(4.3). Note that in Eq.(4.3) we have applied the relation $e_q+e_{q'}=1$
whenever it holds.

  For the case of $q \bar q \rightarrow q \bar q \gamma$ subprocess, we have
\begin{eqnarray}
\vert M(-+) \vert^2 &= &e_q^2R_q^2{{\s^2_{13}+\s^2_
{24}} \over \s_{45}}  \Bigg \{ g_2 \left[{r_{12}^z \over \s_{23}} + {1 \over
 {\s_{12}(\s_{23}-M^2_Z)}}\right] \non \\ &+ & (f_1+g_1)\left[{r_{12}^z \over
\s_{14}}+{1 \over {\s_{12}(\s_{14}-M^2_Z)}}\right]   \non \\
    &  + &(f_2+f_3+g_3)\left[{r_{34}^z \over \s_{23}}
      +{1 \over {\s_{34}(\s_{23}-M^2_Z)}}\right]  \non \\
     & + &(g_1+g_2+g_3)\left[{r_{34}^z \over \s_{14}}
             +{1 \over {\s_{34}(\s_{14}-M^2_Z)}}\right] \Bigg \} \non \\
 & +&  e_q^2R_q^2{{\s^2_{13}-\s^2_{24}} \over \s_{45}}
\Bigg \{ {1 \over D_{12}^z} ({g'_1 \over \s_{14}}
   + {g'_2 \over \s_{23}})
 + {1 \over D_{34}^z} \big({g'_3 \over \s_{23}}
 - {{g'_1+g'_2+g'_3} \over \s_{14}}\big) \Bigg\}, \end{eqnarray}
where $D_{ij}^z = (\s_{ij}-M^2_Z)^2 + M^2_Z \Gamma^2_Z$, $r_{ij}^z=(\s_{ij}-M_Z
^2)/D_{ij}^z$, and
\begin{equation}
g_1' =  {4 \over {\s_{25}\s_{35}}}\ell,\qquad g_2' = {4 \over {\s_{35}
\s_{15}}}\ell, \qquad g_3' ={4 \over {\s_{15}\s_{25}}}\ell,
\end{equation}
with
$\ell\equiv\epsilon_{\mu\nu\alpha\beta} k^{\mu}_1 k^{\nu}_2 k^{\alpha} _3
k^{\beta}_4$, and $|M(+-)|^2$ is obtained from (4.5) by the replacement of
$R_q$ by $L_q$.

For the case of $q \bar q \rightarrow q' \bar q' \gamma$ subprocess,
\begin{eqnarray}
\vert M(-+) \vert^2 &=& -{\s_{13}^2 \over \s_{45} }\Bigg\{ {e_{q'} \over
\s_{12} } (\tilde{A} g_1 + \tilde{B} g_2 +\tilde{C}f_1) \non \\
   &   & + {e_q \over \s_{34}} [ \tilde{A}(f_2+g_3)+\tilde{B}(f_3+g_2)+\tilde
{C}(g_1+g_2) ] \Bigg \} |V_{qq'}|^2 \non \\
   &  &- {\s^2_{24} \over \s_{45} } \Bigg \{ {e_{q'} \over \s_{14}-M^2_{_W}}
       \left[{e_{q'} \over \s_{12}}f_1 + {e_q \over \s_{34} }
   (g_1+g_2)\right]  \non \\  & & + {e_q \over \s_{14} - M^2_{_W} } +
\left[{e_{q'}\over \s_{12} }g_1 +{ e_q \over \s_{34} }(f_2+g_3)\right] \non \\
& & + \left(e_q - {\s_{15} \over \s_{14}-M^2_W}\right){1\over \s_{23}-M^2_W}
 \left[{e_{q'} \over \s_{12} } g_2 + {e_q \over \s_{34}}(g_3+f_3)\right]
                        \Bigg \} |V_{qq'}|^2, \end{eqnarray}
where
\begin{eqnarray}
\tilde{A} &= & {\s_{25} \over {\s_{13}(\s_{23}-M^2_W)}} +
 {(1 + 2 {\s_{25} \over \s_{13}} e_{q'}) \over
  \s_{14} - M^2_W } - {\s_{15}+\s_{45} \over (\s_{14}-M^2_W)(\s_{23}-M^2_W)}
 {\s_{25} \over \s_{13}}, \non \\
\tilde{B}&=&{e_q\over {\s_{23}-M^2_W}}-{\s_{15}\over {(\s_{14} - M^2_W)
(\s_{23}-M^2_W)}},  \non \\
\tilde{C} &= & {e_{q'} \over {\s_{14} - M^2_W}}.  \end{eqnarray}

For the case of $q \bar q' \rightarrow q \bar q'\gamma$ subprocess,
\begin{eqnarray}
\vert M(-+) \vert^2 & = & {\s^2_{13} \over \s_{45}}
                     \Bigg\{ r_{12}^w \bigg[({\s_{25} \over \s_{13} }
   g_3 - g_2){e_qe_{q'} \over \s_{23}}+({  \s_{25} \over \s_{13}}-
                        g_1 e_{q'}){e_{q'} \over \s_{14}}\bigg] \non \\
                      & - &  r_{34}^w \bigg[({\s_{25} \over \s_{13}}+e_{q'})
({g_3e_q\over \s_{23}}+{ g_1 e_{q'} \over \s_{14} }){e_q \over \s_{23}} +
                        ( \s_{35} -{\s_{25} \over \s_{13}}
            (\s_{35}+\s_{45}) {  g_1 e_{q'} \over \s_{14}} \bigg] \non \\
                     &+& r_{12}^w r_{34}^w \bigg[\left(\s_{35} g_2
             -{  \s_{25}(\s_{35} + \s_{45}) \over \s_{13} } g_3\right){
                  e_q \over \s_{23}}+ \left(\s_{35} -{
                   \s_{25} (\s_{35}+\s_{45}) \over \s_{13}}\right){
                         g_1 e_{q'} \over \s_{14}} \bigg] \non \\
    & +& M_W \Gamma_W  \bigg[ (e_{q'}f_1 -{ \s_{25} \over \s_{13}}
                        f_2){ e_{q'} \over \s_{14} D^w_{12}}
               + { e_q^2 f_3 \over \s_{23}D^w_{34}} + ({ \s_{25} \over \s_{13}
} +e_{q'}){e_{q'}f_2 \over \s_{14}}{1 \over \s_{14}D_{34}^w}  \bigg] \non \\
    & + & {M^2_W \Gamma^2_W \over D_{12}^w D_{34}^w}
                          \bigg[ (\s_{35} + \s_{45}){ \s_{25} \over \s_{13}}
                         f_2 - \s_{35} f_1 \bigg] {e_{q'} \over \s_{14}}
                          \Bigg\}|V_{qq'}|^2  \non \\
  & + &{   \s^2_{24} \over \s_{45}}
                         \Bigg\{ r_{12}^w \left({  g_2 e_q \over \s_{23} }+{
            g_1 e_{q'} \over \s_{14} } \right) e_{q'}
    + r_{34}^w \left( { g_3 e_qe_{q'} \over \s_{23}
                        } + {g_1 e_{q'}^2 \over \s_{14}}- {
         g_2 e_qe_{q'} \over \s_{14} }-{g_3 e_qe_{q'} \over \s_{14}
                        }\right)  \non \\
 &+& r_{12}^w r_{34}^w \left( f_1e_{q'} {
                                \s_{35} \over \s_{14}
                        } -g_1e_{q'} {\s_{35}\over \s_{14}
                        }- g_2 e_q {\s_{35} \over \s_{23}  }\right)\non  \\
  & - & M_W \Gamma_W \left({f_1 e_{q'}^2 \over D_{12}^w\s_{14}
                        } + { f_2 e_{q'}^2 \over D_{34}^w \s_{14} } + {
            f_3 e_{q}^2 \over D_{34}^w\s_{23}  }\right)  \non \\
& + & { M^2_W \Gamma^2_W \over D_{12}^w D_{34}^w } (g_1e_{q'}+g_2 e_q -
f_1e_{q'}) {\s_{35} \over \s_{14}  }\Bigg\}|V_{qq'}|^2 \non \\
& + & N(-+),   \end{eqnarray}
where $N(-+)$ involves the totally antisymmetric tensor $\epsilon_{\mu\nu\rho
\sigma}$ and reads
\begin{eqnarray}
 {\s^2_{13} \over \s_{45}} & \Bigg\{ & \bigg\{ \bigg[
     ( g'_2 - {\s_{25} \over \s_{13}} g'_3 ) {e_qe_{q'}\over \s_{23}}
+ {1\over \s_{14}} ( g'_1e_{q'}^2 + g'_1e_{q'}{\s_{25}\over \s_{13}} ) \bigg]
     {1 \over D^w_{12}}  + \bigg[ (
     {\s_{25} \over \s_{13}} + e_{q'} ) ( {g'_3 e_q \over \s_{23}} -
     {g'_1 e_{q'} \over \s_{14}}) \non \\
   & & - (g'_2 +g'_3) {e_qe_{q'} \over \s_{14}} \bigg]
     {1 \over D_{34}^w} \bigg\}  M_W \Gamma_W
      + {M^2_W \Gamma^2_W \over D_{12}^w D_{34}^w} \bigg[
 \left(g'_3(\s_{35}+\s_{45}) {\s_{25} \over\s_{13}}-g'_2\s_{35}\right)
     { e_q \over \s_{23}} \non \\
   &  & + \left((\s_{35}+\s_{45}){\s_{25} \over\s_{13}}-\s_{35}\right){g'_1
e_{q'}\over \s_{14}}\bigg]\Bigg\}|V_{qq'}|^2  \non \\
     + {\s^2_{24} \over \s_{45}}
  &   \Bigg\{ & M_W \Gamma_W   \bigg[
     {e_{q'} \over D_{12}^w} ( {g'_2 e_q \over \s_{23}} +
     {g'_1 e_{q'} \over \s_{14}} )    + {e_{q'} \over D_{34}^w}
     ( {g'_3 e_q \over \s_{23}} - {g'_1 e_{q'}\over \s_{14}}
     + {g'_2 e_q \over \s_{14}} + {g'_3 e_q \over \s_{14}} )  \non \\
   &  & + ( {r_{12}^w \over D_{34}^w} + {r_{34}^w \over D_{12}^w}) (
 {g'_2 e_q \over \s_{23}} + {g'_1 e_{q'} \over \s_{14}} ) \s_{35}  \bigg]
\Bigg\}|V_{qq'}|^2, \end{eqnarray}
with $D^w_{ij} = (\s_{ij} - M_W)^2 + M_W^2 \Gamma_W^2 $ and
$r^w_{ij} = (\s_{ij} - M_W^2)/ D^w_{ij}$.

   Finally, the numerator of the single asymmetry $\a$ for 2-jet plus photon
production is given by
\begin{eqnarray}
 \sigma^+-\sigma^- & = & {32\over 9}\,{\alpha^2 \alpha_s\over s_{12} x_W (1-
x_W)}\sum_{i,j} \int dx_adx_b\Delta f_i^{(a)}(x_a,Q^2)f_j^{(b)}(x_b,Q^2)
\non \\  & \times & \vert\Delta M \vert^2
         {d^3k_3 \over E_3} {d^3k_4 \over E_4} {d^3k_5 \over E_5}
\delta^4 (k_1 + k_2 - k_3  - k_4 - k_5),
\end{eqnarray}
where $x_W\equiv\sin^2\theta_W$, $\vert\Delta M\vert^2={1\over 2}(\vert M(++)
\vert^2+\vert M(+-)\vert^2-\vert M(-+)\vert^2-\vert M(--)\vert^2)$.
The unpolarized cross sections for 2-jet plus
photon production can be found in Ref.[3].

\vskip 0.6 cm
\noindent {\bf V.~~$\ell^+ \ell^-$ pair plus 1-jet production}
\vskip 0.4 cm
  Since QCD interactions change color whereas weak couplings always conserve
color, pv effects at tree level in general depend only on the polarized
 valence and sea quark distributions. It is thus desirable to have some
processes in
which gluons also contribute to the pv asymmetry $\a$. In this section we
shall study one of such
reactions, namely $pp\to \ell^+\ell^-+{\rm jet}+X$.
\footnote{The pv effect in the reaction $pp\to\ell^+\ell^-+X$ has been
discussed in Ref.[22].}

At the parton level, there are two subprocesses contributing to the Drell-Yan
reaction $pp\to \ell^+\ell^-+{\rm jet}+X$: $G+q(\bar{q})\to \ell^+\ell^-+q(\bar
{q})$ and $q\bar{q}\to\ell^+\ell^-+G$. In this reaction, $\a$ arises from the
interference between the amplitudes with $\gamma$ and $Z^0$ exchanges.
The transition matrix elements absolute squared for the
subprocess $G_{\lambda_1}(k_1) + q_{\lambda_2}(k_2) [\bar q_{\lambda_2}(k_2)]
\rightarrow \ell^+(k_3) \ell^-(k_4) + q(k_5) [\bar q(k_5)]$
are (only the interference terms being written down)
\renewcommand{\theequation}{5.\arabic{equation}}
\setcounter{equation}{0}
\begin{eqnarray}
\vert M(++) \vert ^2 &= & 2 (4 \pi \alpha_s) (4 \pi \alpha)^2 { \s_{34} \over
                   \s_{12} \s_{15} } \Bigg\{
                    { R_q^2 (L_{\ell}^2 \s^2_{24} + R_{\ell}^2 \s^2_{23}) \over
                    4 x^2_W (1-x_W)^2 D^z_{34} } \non \\
                  & + & 4 {e_q^2 \over \s^2_{34} } +
                 2 R_q r^z_{34}e_q{ (L_{\ell} \s^2_{24} + R_{\ell} \s^2_{23})
                    \over x_W (1-x_W) \s_{34} }\Bigg\}, \end{eqnarray}
\begin{eqnarray}
\vert M(+-) \vert ^2 &= & 2 (4 \pi \alpha_s) (4 \pi \alpha)^2 { \s_{34} \over
                   \s_{12} \s_{15} } \Bigg\{
                    { L_q^2 (L_{\ell}^2 \s^2_{45} + R_{\ell}^2 \s^2_{35}) \over
                    4 x^2_W (1-x_W)^2 D^z_{34} } \non \\
                &  +  & 4{ e_q^2 \over \s^2_{34} } +
                 2 L_q r^z_{34}e_q{ (L_{\ell} \s^2_{45} + R_{\ell} \s^2_{35})
                    \over x_W (1-x_W) \s_{34} } \Bigg\},  \end{eqnarray}
where
\begin{eqnarray}
R_\ell=2\sin^2\theta_W,~~~~L_\ell=-1+2\sin^2\theta_W.
\end{eqnarray}
Using the hermitian conjugate of $M(++)$ and $M(+-)$, we obtain
\begin{eqnarray}
\vert M(--) \vert ^2& = & 2 (4 \pi \alpha_s) (4 \pi \alpha)^2 { \s_{34} \over
                   \s_{12} \s_{15} } \Bigg\{
                    { L_q^2 (L_{\ell}^2 \s^2_{24} + R_{\ell}^2 \s^2_{23}) \over
                    4 x^2_W (1-x_W)^2 D^z_{34} } \non \\
                &   + & 4{ e_q^2 \over \s^2_{34} } +
                 2 L_q r^z_{34}e_q{ (L_{\ell} \s^2_{24} + R_{\ell} \s^2_{23})
                    \over x_W (1-x_W) \s_{34} } \Bigg\},  \end{eqnarray}
and
\begin{eqnarray}
\vert M(-+) \vert ^2 & = & 2 (4 \pi \alpha_s) (4 \pi \alpha)^2 { \s_{34} \over
                   \s_{12} \s_{15} } \Bigg\{
                    { R_q^2 (L_{\ell}^2 \s^2_{45} + R_{\ell}^2 \s^2_{35}) \over
                    4 x^2_W (1-x_W)^2 D^z_{34} } \non \\
                 &  + & 4{ e_q^2 \over \s^2_{34} } +
                 2 R_q r^z_{34}e_q{ (L_{\ell} \s^2_{45} + R_{\ell} \s^2_{35})
                    \over x_W (1-x_W) \s_{34} }  \Bigg\}. \end{eqnarray}

 The  transition matrix elements absolute squared for the subprocess of
$$q_{\lambda_1}(k_1) + \bar q_{\lambda_2}(k_2) \rightarrow
\ell^+(k_3) \ell^-(k_4) + G(k_5) $$
read
\begin{eqnarray}
\vert M(-+) \vert ^2 & = & 2 (4 \pi \alpha_s) (4 \pi \alpha)^2 { \s_{34} \over
                   \s_{15} \s_{25} } \Bigg\{
                    { L_q^2 [ R_{\ell}^2 (\s^2_{14}  + \s^2_{24})
          + L_{\ell}^2 ( \s^2_{13} + \s^2_{23}) ] \over
                    4 x^2_W (1-x_W)^2 D^z_{34} } \non \\
                 &  + & 4{ e_q^2 \over \s^2_{34} } +
            2 L_q r^z_{34}e_q{[ L_{\ell} (\s^2_{14} + \s^2_{24})
                    + R_{\ell} ( \s^2_{13} +\s^2_{23})]
                    \over x_W (1-x_W) \s_{34} }
              \Bigg\}, \end{eqnarray}
and
\begin{eqnarray}
\vert M(+-) \vert ^2 & = & 2 (4 \pi \alpha_s) (4 \pi \alpha)^2 { \s_{34} \over
                   \s_{15} \s_{25} }  \Bigg\{
                    { R_q^2 [ R_{\ell}^2 (\s^2_{14}  + \s^2_{24})
          + L_{\ell}^2 ( \s^2_{13} + \s^2_{23}) ] \over
                    4 x^2_W (1-x_W)^2 D^z_{34} } \non \\
                  & + & 4{ e_q^2 \over \s^2_{34} } +
            2 R_q r^z_{34}e_q{ [ L_{\ell} (\s^2_{14} + \s^2_{24})
                    + R_{\ell} ( \s^2_{13} +\s^2_{23}) ]
                    \over x_W (1-x_W) \s_{34} }
              \Bigg\}. \end{eqnarray}

\pagebreak
\noindent{\bf VI.~~Results and discussion}
\vskip 0.4 cm
  The results of our calculations for parity-violating asymmetries $\a$ and
$\p$ are presented in a series of figures. Shown in Figs.1 and 2 are the
$\sqrt{s}$ dependence of the helicity asymmetric cross section $\Delta\sigma_L
=\sigma^+
-\sigma^-$ and $\a$, respectively, at RHIC energies for a single jet production
at $90^\circ$ (i.e. $y=0$) in the c.m. with a jet momentum cutoff at 5 GeV for
three different cases of polarized parton distributions (see Sec.II).
\footnote{For the valence $d$ quark distribution function, we use $x_0=0.50$
in realistic calculation [see Eq.(2.9)].}
Note that all parton spin densities constrained by the EMC data
are first parametrized at $Q_0^2=10\,{\rm
GeV}^2$ [Eqs.(2.11-2.13)] and then their $Q^2$ evoluations are governed by the
Altarelli-Parisi equations. Since at tree level gluons in general do not
contribute to $\a$ and $\p$, pv asymmetries are only sensitive to polarized
valence and sea quark distribution functions. From Figs.1 and 2 we see that
$\a$ is of order $10^{-5}$ at RHIC energies and is dominated by
case (i) with large sea polarization. This is what expected since as
far as $\a$ is concerned, the three different parametrizations of parton spin
densities are different only in their sea polarization. The behavior of
$\a$ in the 2-jet production at rapidities $y_1=y_2=0$ with the jet momentum
being cut off at 5 GeV has the same pattern
as the previous 1-jet case (see Figs.1-4). This is
attributed to that the underlying parton-parton scatterings for
1-jet and 2-jet productions are the same.

   We have calculated in Figs.5 and 6 the longitudinal polarization $\p$ of
$\Lambda$ produced at $y=0$ in unpolarized $pp$ collisions at energies
$\sqrt{s}=200$ GeV and 500 GeV as a function of $x_T=2p_T/\sqrt{s}$. The
dependence of $\p$ on $x_T$, which is quite similar at the two energies shown
in Fig.6, is one of the testable predictions for the polarized $\Lambda$
fragmentation
function given in Eq.(2.17). The increase of $\p$ with $x_T$ basically can be
understood from the $z$ dependence of $\Delta D^\Lambda_q(z)/D^\Lambda_q(z)=z
^\gamma$ with $\gamma>0$; especially at large $x_T$ where the unpolarized
cross section is dominated, as noted in passing, by quark-quark scattering.
In order to have a numerical estimate for $\p$, we have followed Ref.[11] to
choose $\gamma=10$. The resulting $\p$ is of order $10^{-2}$ at moderate
$x_T$. In practice, it will be easier to measure the $\Lambda$
polarization at small $x_T$ where the signal of $\Delta\sigma_\Lambda$ is
large. As stressed in Sec.II, very little is known about the fragmentation
functions
$D^\Lambda_q(z)$ and $\Delta D^\Lambda_q(z)$. Presumably, the experiment of
searching for helicity asymmetric cross section  in semi-inclusive production
of $\Lambda$ in $e^+e^-$ annihilation allows one to measure the polarized
fragmentation function $\Delta D^\Lambda_q(z)$ [21].

   We have also extended our discussions to the hadronic production of 2-jet
plus photon final states. We plot in Figs.7 and 8 the differential cross
section asymmetry and $\a$, respectively, as a function of the invariant mass
$\sqrt{s_{34}}$ of
two jets. Since experimentally it is difficult to distinguish between quark
and antiquark jets, we have symmetrized these two jets. The predicted $\a$
increases from $10^{-4}$ to $10^{-3}$ as $\sqrt{s}$ varies from 100 GeV to 500
GeV, and it is insensitive to the polarized structure functions chosen.

  Our investigation of the Drell-Yan type reaction $pp\to\ell^+\ell^-+{\rm
jet}+X$
is originally motivated by looking for the processes in which gluons also
contribute to pv asymmetry so that a measurement of $\a$ in such reactions
would provide useful information on the gluon polarization. The results are
presented in Figs.9 and 10. It is evident from Fig.9 that when the invariant
mass of dilepton is around the mass of the $Z^0$ resonance,
a bump is shown up in both unpolarized and helicity asymmetric
cross sections, as it should be. Unfortunately, the resulting $\a$
 is insensitive to the choice of parton spin densities. This may be ascribed
to the fact that contributions to the helicity asymmetric cross section
come mainly from the region of moderate
value of $x$ where spin dependent gluon and sea distributions are negligible.
(Also note that for the subprocess $G+q\to\ell^+\ell^-+q$, $|\Delta M|^2$
arising from polarized gluons has a sign opposite to that due to polarized
quarks [see Eqs.(5.1)-(5.5)].) We conclude that, contrary to the double
helicity helicity asymmetry ${\cal A}_{LL}$, it is unlikely that a
measurement of the single helicity asymmetry $\a$ will bring any new insight
into the gluon polarization.

   Finally, we note that a large pv asymmetry $\a$ of order
$10\%$ is expected to be seen in $W^\pm$ and $Z^0$ productions with a large
$p_T$ at RHIC energies (see
Ref.[22] and the first paper in Ref.[1] for details). On the other hand, the
helicity asymmetric cross section is estimated to be of order $1pb$ for $W^+$
(and even smaller for $W^-$ and $Z^0$) in $pp$ collisions at energy, say
$\sqrt{s}=500$ GeV, to be compared with $\Delta\sigma\sim
10^{-1}nb$ for 1-jet or 2-jet production (see Figs.1 and 3). Therefore, it is
worth pursuing all possible parity-violating effects in high energy
hadron-hadron collisions at the planned hadronic colliders RHIC, SSC and LHC.

\pagebreak

\centerline{\bf Acknowledgments}
\vskip 0.8 cm
One of us (H.Y.C.) wishes to thank Prof. C. N. Yang and the Institute for
Theoretical Physics at Stony Brook for their hospitality during his stay
there for sabbatical leave. C.F.W. would like to
thank Profs. Gonsalves and C. Y. Cheung for many helpful discussions.
This work was supported in part by  the National Science
Council of the Republic of China under Contract Nos.
NSC82-0208-M001-001Y and  NSC82-0112-C001-019.

\vskip 2.5 cm
\centerline {\bf REFERENCES}
\vskip 1.0cm
\begin{enumerate}

\item C. Bourrely, J. Ph. Guillet, and J. Soffer, \np {\bf B361}, 72 (1991);
H.Y. Cheng, S. R. Hwang, and S.N. Lai, \pr {\bf D42}, 2243 (1990).
\item D. Indumathi, M.V.N. Murthy, and V. Ravindran, IMSc-91/31 (1991).
\item M. A. Doncheski, R. W. Robinett and L. Weinkauf, \pr {\bf
D44}, 2717 (1991).
\item C. F. Wai, IP-ASTP-15-1992 (submitted to Phys. Lett. B).
\item M. A. Doncheski and R. W. Robinett, \pr {\bf D46}, 2011 (1992).
\item E. Berger and J. Qiu, \pr {\bf D40}, 778 (1989); S. Gupta, D,
Indumathi and M. V. N. Murthy {\sl Z. Phys.} {\bf C42}, 493 (1989); {\sl ibid.}
 {\bf C47}, 227 (1990); K. Kobayakawa, T. Morii, and T. Yamanishi,
KOBE-FHD-92-03 (1992).
\item H. Y. Cheng and S. N. Lai, \pr {\bf D41}, 91 (1990).
\item The EMC Collaboration, J. Ashman {\it et al.}, \pl {\bf B206}, 264
(1988); \np {\bf B328}, 1 (1989).
\item R.L. Jaffe and A. Manohar, \np {\bf B337}, 509 (1990).
\item A.V. Efremov and O.V. Teryaev, in {\it Proceedings of the International
Hadron Symposium}, Bechynue, Czechoslovakia, 1988, edited by X. Fischer
{\it et al.} (Czechoslovakian Academy of Science, Prague, 1989), p.302;
G. Altarelli and G.G. Ross, \pl {\bf B212}, 391 (1988);
R.D. Carlitz, J.C. Collins, and A.H. Mueller, \pl {\bf B214}, 229 (1988).
\item H.Y. Cheng and E. Fischbach, \pr {\bf D19}, 2123 (1979).
\item M. Abud, R. Gatto, and C.A. Savoy, {\sl Ann. Phys. (N.Y.)} {\bf 122},
219 (1979).
\item G. Ranft and J. Ranft, \pl {\bf B87}, 122 (1979); \np {\bf B165}, 395
(1980).
\item H.Y. Cheng and C.F. Wai, \pr {\bf D46}, 125 (1992).
\item M. Diemoz, F. Ferroni, E. Longo, and G. Martinelli, {\sl Z. Phys.} {\bf
C39}, 21 (1988).
\item G. R. Farrar and D. R. Jackson, {\sl Phys. Rev. Lett.} {\bf 35}, 1416
(1975).
\item G.T. Bodwin and J.W. Qiu, \pr {\bf D41}, 2755 (1990); and talk presented
by Bodwin at the Polarized Collider Workshop, Penn State Univ. November 1990.
\item We employ the numerical analysis program written by W.K. Tung.
\item R. Badier {\it et al.}, {\sl Z. Phys.} {\bf C2}, 265 (1979).
\item The EMC Collaboration, M. Arneodo {\it et al.,} \np {\bf B321}, 541
(1989).
\item M. Burkardt and R.L. Jaffe, {\sl Phys. Rev. Lett.} {\bf 70}, 2537 (1993).
\item See e.g., C. Bourrely, J. Soffer, F. M. Renard, and P. Taxil, {\sl Phys.
Rep.} {\bf 177}, 319 (1989).
\item R.D. Field, {\it Application of Perturbative QCD} (Addison-Wesley, 1989).

\end{enumerate}
\vfill\eject

\centerline{\bf FIGURE CAPTIONS}
\vskip 1cm
\begin{enumerate}
\item The signal of helicity asymmetric cross section for 1-jet production at
$y=0$ in $pp$ collisions as a function of $\sqrt{s}$ for three different
polarized parton distributions as described in the text.
\item Asymmetry $\a$ for 1-jet production at $y=0$ in $pp$ collisions as a
function of $\sqrt{s}$ for three different polarized parton distributions as
described in the text.
\item Same as Fig.1 except for 2-jet production at rapidities $y_1=y_2=0$.
\item Same as Fig.2 except for 2-jet production at rapidities $y_1=y_2=0$.
\item The signal of helicity asymmetric cross section for $\Lambda$ production
at $y=0$ in unpolarized $pp$ collisions as a function of $x_T$ at energies
$\sqrt{s}=200$ and 500 GeV.
\item Longitudinal polarization $\p$ of  $\Lambda$ produced at $y=0$ in
unpolarized $pp$ collisions as a function of $x_T$ at energies $\sqrt{s}=200$
and 500 GeV.
\item The signal of helicity asymmetric cross section  for 2-jet plus photon
production
in $pp$ collisions as a function of the invariant mass $\sqrt{s_{34}}$ of
dijet for three different polarized parton distributions.
\item Asymmetry $\a$ for 2-jet plus photon production in $pp$ collisions
as a function of the invariant mass $\sqrt{s_{34}}$ of dijet for three
different polarized parton distributions.
\item The signal of helicity asymmetric cross section for $\ell^+\ell^-$ plus
1-jet production in $pp$ collisions as a function of the invariant mass
$\sqrt{s_{34}}$ of dilepton for three different polarized parton distributions.
\item Asymmetry $\a$ for $\ell^+\ell^-$ plus 1-jet production in $pp$
collisions as a function of the invariant mass $\sqrt{s_{34}}$ of dilepton
for three different polarized parton distributions.
\end{enumerate}

\end{document}